\documentclass[prl,aps,twocolumn,groupedaddress,showpacs,floatfix,preprintnumbers]{revtex4}
\usepackage{amsmath,amssymb,multirow,epsfig,bm}

\newcommand{\avg}[1]{\langle #1 \rangle}

\newcommand{\fig}[1]{Fig.~\ref{#1}}
\newcommand{\apriori}{\textit{a~priori} }              % a priori
\newcommand{\pictsize}{1.0}
\newcommand{\beq}{\begin{equation}}
\newcommand{\eeq}{\end{equation}}
\newcommand{\bea}{\begin{eqnarray}}
\newcommand{\eea}{\end{eqnarray}}
\newcommand{\pcut}{k^{}_{\text{max}}}  
 
\newcommand{\eF}{\varepsilon_{F}}
\newcommand{\pF}{k_{F}}
\newcommand{\reff}{r^{}_{\textrm{eff}}}

\begin{document}
\preprint{NT@UW-15-01}

\title{Perfect-fluid behavior of a dilute Fermi gas near unitary}

\author{  Gabriel Wlaz\l{}owski$^{1,2}$, Wei Quan$^{2}$, and Aurel Bulgac$^{2}$ }

\affiliation{$^1$Faculty of Physics, Warsaw University of Technology,
  Ulica Koszykowa 75, 00-662 Warsaw, Poland}
\affiliation{$^2$Department of Physics, University of Washington,
  Seattle, Washington 98195--1560, USA}
 
\begin{abstract}
  We present an {\it ab initio} calculation of the shear viscosity as
  a function of interaction strength in a two-component unpolarized
  Fermi gas near the unitary limit, within a finite temperature
  quantum Monte Carlo (QMC) framework and using the Kubo
  linear-response formalism. The shear viscosity decreases as we tune
  the interaction strength $\frac{1}{a\pF}$ from the
  Bardeen-Cooper-Schrieffer side of the Feshbach resonance
  towards Bose-Einstein condensation limit and it acquires the
  smallest value for $\frac{1}{a\pF}\approx0.4$, with a minimum value
  of
  $\left . \frac{\eta}{s}\right |_{min} \approx 0.2
  \frac{\hbar}{k_B}$,
  which is about twice as small as the value reported for
  experiments in quark-gluon plasma
  $\left . \frac{\eta}{s}\right |_{QGP} \lesssim
  0.4\frac{\hbar}{k_B}$.
  The Fermi gas near unitarity thus emerges as the most ``perfect
  fluid" observed so far in nature. The clouds of dilute Fermi gas
  near unitarity exhibit the unusual attribute that, for the sizes
  realized so far in the laboratory or larger (less than $10^9$ atoms),
  can sustain quantum turbulence below the critical temperature,
  but at the same time the classical turbulence is suppressed in 
  the normal phase.
  
\end{abstract}

\date{\today}

% 03.75.Ss Degenerate Fermi gases
% 05.30.Fk Fermion systems and electron gas
% 03.75.Hh Static properties of condensates; thermodynamical,statistical,and structural properties
% 67.25.dt Sound and excitations
% 67.85.Lm Degenerate Fermi gases
% 05.60.Gg Quantum transport
% 51.20.+d Viscosity, diffusion, and thermal conductivity

\maketitle

%%%%%%%%%%%%%%%%%%%%%%%%%%%%%%%%%%%%%%%%%%%%%%%%%%%%%%%%%%%%%%
%-------------------------------------------------------------
% ------------------------ INTRODUCTION ----------------------
%-------------------------------------------------------------

\section{I. INTRODUCTION}

The ultracold atoms provide an ideal laboratory for very precise
experimental and theoretical studies of an enormous range of quantum
mechanical phenomena. A large set of studies focused on interacting
Fermi gases at unitarity, where such systems exhibit remarkable
properties. The average inter-particle separation $\approx n^{-1/3}$
is large compared to the effective range of interaction $\reff$, but
small compared to the scattering length $|a|$, i.e.,
$0 \leftarrow n^{1/3}\reff\ll 1 \ll n^{1/3}|a|\rightarrow\infty$
(where $n$ is particle density). In this limit these systems acquire
universal properties and they have been widely studied over the last dozen
years both experimentally and theoretically (see
reviews~\cite{reviews1,reviews2,bcsbec,FirstExperiments}).  A series
of recent experiments~\cite{Turlapov,Cao1,Cao2,Joseph} revealed the
nearly ideal hydrodynamic behavior of the resonantly interacting Fermi
gas, characterized by a very low shear viscosity coefficient to the
entropy density ratio - very close to the conjectured bound
originating from holographic duality methods~\cite{KSS,holographic},
called KSS bound. An ``ideal fluid'' which follows the laws of ideal
hydrodynamics, is a fluid in which dissipative processes are
absent~\cite{LandauLifshitz}.  The superfluid component of a quantum
fluid below the critical temperature is treated as a physical
realization of an ideal fluid~\cite{Khalatnikov}.  The normal component is
characterized by a finite viscosity, which in an infinite medium tends
to infinity at absolute zero temperature, due to the contributions of
phonons~\cite{Khalatnikov}. In a finite system however the longest
phonon wavelength is of the order of the size of the container and
with the number of excited phonons $\propto T^3$ this has the 
consequence that the viscosity, while approaching the zero 
temperature, never formally diverges and a liquid is effectively in 
a collisionless regime.

A hypothetical physical system which saturates the KSS bound and has
the lowest possible value of the shear viscosity is often referred to
as ``the perfect fluid''.  A recent measurement of the shear viscosity
of a unitary gas close to unitarity~\cite{Joseph} provides the value
$\frac{\eta}{s}\approx 0.5\frac{\hbar}{k_B}$, while KSS bound is
$\frac{1}{4\pi}\frac{\hbar}{k_B}\approx 0.08 \frac{\hbar}{k_B}$.  Thus
the unitary Fermi gas (UFG) appears as the system being very close to
``perfectness.'' Another system is the quark-gluon plasma created in
relativistic heavy ion collisions~\cite{Teaney_2009}, where the value
$\left . \frac{\eta}{s}\right |_{QGP} \lesssim 0.4\frac{\hbar}{k_B}$
has been reported~\cite{Song}.

\begin{figure}
\includegraphics[width=\pictsize\columnwidth]{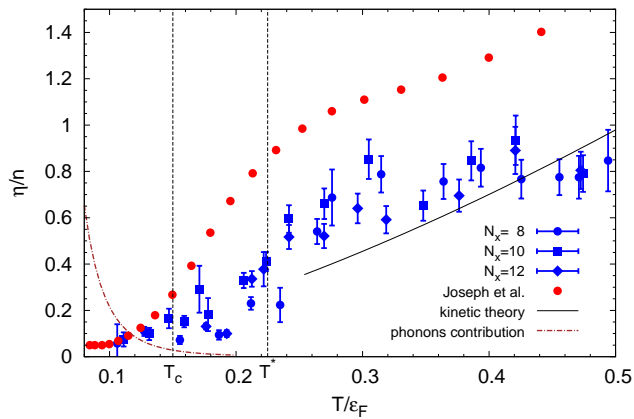}
\caption{ (Color online) Comparison of the QMC predictions taken from
Ref.~\cite{Wlazlowskietal3} (blue points) with the results extracted
from experimental data for trapped system (red
points)~\cite{Joseph}. Horizontal and vertical axes show the
temperature in units of Fermi energy $\eF=\frac{\hbar^2\pF^2}{2m}$ and the
dimensionless shear viscosity $\eta/n$, respectively.  The theoretical
results are provided for different lattice sizes $N_x=8,10$ and
$12$. The scatter of these points is a measure of the numerical
accuracy of the available QMC predictions. The solid black line shows kinetic theory 
prediction $\frac{\eta}{n} \cong 2.77 \left (\frac{T}{\eF}\right )^{3/2}.$
The phonon contribution to the viscosity, which is not accounted for in 
QMC calculations in a finite volume~\cite{Wlazlowskietal3} and evaluated 
in Ref.~\cite{RupakSchafer}, is shown as a dot-dashed (brown) line.
The vertical black dotted lines indicate
the critical temperature of superfluid-to-normal phase transition for
$T_c=0.15(1)\, \eF$ and the onset of Cooper pair formation,
$T^*\approx0.22\,\eF$, respectively - both extracted from QMC
simulations in Refs~\cite{BDM} and~\cite{Wlazlowskietal2}.
\label{fig:expcmp} }
\end{figure}

Simultaneously, an impressive effort has been made in order to
theoretically determine transport coefficients of the UFG. A plethora
of theoretical methods has been used to estimate these transport
coefficients, both for homogeneous and trapped systems~
\cite{BruunSmith,RupakSchafer,Schafer,Enssetal,Guoetal,
  Brabyetal,SalasnichToigo,LeClair,Wlazlowskietal,Wlazlowskietal3}.
Typically these theoretical predictions differ both quantitatively and
qualitatively. Among the methods used, only the
works~\cite{Wlazlowskietal,Wlazlowskietal3} present \textit{ab initio}
calculations, obtained within the very powerful quantum Monte Carlo
(QMC) framework, where the errors can be quantified quite
accurately. In these papers, however, the shear viscosity has been
evaluated only at unitarity.  In \fig{fig:expcmp} we compare these
calculations with recent experimental data for a uniform system
extracted from measurements for trapped
systems~\cite{Joseph}. Agreement for absolute value of shear viscosity
has not been obtained, which however does not rule out the QMC
results. However, we observe a very good qualitative
agreement. Experimentally, the results for uniform
systems~\cite{Joseph} have been inferred from data for trapped systems
under a number of assumptions (see Supplement of Ref.~\cite{Joseph}). To
what extent these assumptions are valid, especially in the most
interesting low temperature regime, is not clear. In particular,
viscous hydrodynamics has been used in Ref.~\cite{Joseph} to describe
the full dynamics of the cloud, even at the periphery of the cloud,
where the density is very low and the collision rate is strongly
suppressed and the system is in the collisionless
regime~\cite{Urban_2015}.  In both cases we observe the rapid decrease
of the shear viscosity well above critical temperature of
superfluid-normal phase transition $T_c$. It is notable that this
feature is present only in the QMC and the pseudogap $T$-matrix
theory~\cite{Guoetal} predictions.

%%%%%%%%%%%%%%%%%%%%%%%%%%%%%%
%%%%%%%%%%%%%%%%%%%%%%%%%%%%%%

Recently it has been reported that the system is even closer to the
prefect fluid limit if it is slightly beyond the unitary
point~\cite{Elliott}. In this experiment it was revealed that the
lowest viscosity coefficient can be obtained for an interaction
strength corresponding to $\frac{1}{a\pF}\approx0.25$, where
$\pF=(3\pi^2 n)^{1/3}$ is Fermi momentum. Since the QMC method
provides results which are at least in very good qualitative agreement
with experiment for the unitary limit (i.e., $\frac{1}{a\pF}\to 0$), in
this work we extend our studies beyond unitarity point.

\section{II. METHOD}

In order to determine the shear viscosity coefficient of the ultracold
atomic gas we employ the QMC technique with auxiliary fields on the
lattice, which provides numerical results with controllable accuracy,
up to quantifiable systematic uncertainties (for details see
Ref.~\cite{BDM}).  These simulations are very similar to those
performed at unitarity when $|a|=\infty$
(Refs.~\cite{Wlazlowskietal,Wlazlowskietal3}).  Here we briefly describe
the main aspects of the computational process, focusing mainly on
required modifications in order to study the system beyond the
unitarity. Henceforth we use the system of units: $\hbar=m=k_{B}=1$.

For simulations we employed a cubic lattice of size $N_x=N_y=N_z=10$,
with lattice spacing $l=1$ and average number densities
$n\simeq 0.04$. As shown in the Supplemental Material of
Refs.~\cite{Wlazlowskietal,Wlazlowskietal2} the systematic errors are
no more than $10\%$ for this lattice size and the errors are related
mainly to corrections coming from the nonzero effective range $\reff$
and from the exclusion of the universal high momenta tail in the
occupation probability due to the finite momentum cut-off
$\pcut=\frac{\pi}{l}$.  In the
context of computation of transport coefficient it is important to
have as small as possible statistical errors, so as to minimize the
errors arising from the analytical continuation from the imaginary
time to real frequencies as explained below. Thus we generate an
ensemble containing about $10^4$ uncorrelated samples in order to get
statistical accuracy below $1\%$.

In order to extract the shear viscosity within a QMC framework one
calculates the imaginary-time (Euclidean) stress tensor-stress tensor
correlator
%%%%%%%%%%%%%%%%%%%%%%%%%%%%%%
\begin{equation}
 G_{\Pi}(\bm{q},\tau)=\dfrac{1}{V}
 \avg{\hat{\Pi}_{\bm{q}}^{(xy)}(\tau)\hat{\Pi}_{-\bm{q}}^{(xy)}(0)},
\label{eqn:GPi}
\end{equation} 
%%%%%%%%%%%%%%%%%%%%%%%%%%%%%%
at zero momentum $\bm{q}=0$. As shown in Ref.~\cite{Enssetal}, for
zero-range interaction it is sufficient to use only the kinetic part
of the stress tensor:
%%%%%%%%%%%%%%%%%%%%%%%%%%%%%%
\begin{equation}
\hat{\Pi}_{\bm{q}=0}^{(xy)}=\sum_{\bm{p},\lambda=\uparrow,\downarrow}
p_x p_y\hat a^\dagger_\lambda(\bm{p}) \, \hat a_\lambda^{}(\bm{p}).
\end{equation}
%%%%%%%%%%%%%%%%%%%%%%%%%%%%%%
The average is performed in the grand canonical ensemble, at fixed
temperature $T=\frac{1}{\beta}$ and chemical potential,
${\hat{\Pi}}_{ \bm{q} }^{ (xy) }(\tau) = \exp [
\tau(\hat{H}-\mu\hat{N})] {\hat{\Pi}}_{ \bm{q} }^{ (xy) } \exp
[-\tau(\hat{H}-\mu\hat{N})] $,
where $\hat{H}$ is the Hamiltonian of the system, $\mu$ is the
chemical potential, and $\hat{N}$ is the particle number operator.  In
order to capture the physics of a dilute fermionic gas it is
sufficient to use a zero-range two-body interaction
$V(\bm{r}_1-\bm{r}_2)=-g\delta(\bm{r}_1-\bm{r}_2)$, where the coupling
constant $g$ can be tuned to fix the value of the $s$-wave scattering
length $a$~\cite{BDM} using a standard renormalization procedure of
the coupling constant on the lattice.  After this procedure the
contact interaction acquires a finite effective range - for the
parameters of presented simulation $\pF\reff\simeq 0.43$.  In this
case the interaction part of the stress tensor also contributes to the
correlator.  However, we relegate all the corrections arising from the
finite effective range to the systematic errors, which were estimated
to be less than $10\%$ for used lattice (see the Supplemental Material of
Ref.~\cite{Wlazlowskietal2} for extensive discussion in the context of
lattice computation of transport coefficients).  It is legitimate to
drop this contribution, particularly since in the limit
$\pF\reff \rightarrow 0$ it is vanishing anyway, a fact also
consistent with previous QMC studies performed at unitarity for
different densities.  We emphasize that the dominant source of
uncertainties is introduced by the analytic continuation procedure
(described next).  These are of the order $10\%$, as one can judge
from \fig{fig:expcmp}, where the error bars show contribution to the
uncertainties only from this source.

The frequency dependent shear viscosity $\eta(\omega)$ is extracted
via the analytic continuation of the imaginary-time correlator to real
frequencies. This procedure is equivalent to solving the integral
equation:
%%%%%%%%%%%%%%%%%%%%%%%%%%%%%%
\begin{equation}
  G_{\Pi}(\bm{q}=0,\tau)=\dfrac{1}{\pi}
  \int_{0}^{\infty}\eta(\omega)\,
  \omega\dfrac{\cosh\left[ \omega(\tau-\beta/2)\right] }{\sinh\left( \omega\beta/2\right) }
  d\omega .
  \label{eqn:InversionProblem}
\end{equation}
%%%%%%%%%%%%%%%%%%%%%%%%%%%%%%
The static shear viscosity $\eta$ is defined as
$\eta=\lim_{\omega\to 0}\eta(\omega)$.  The correlator $ G_{\Pi}$ is
sampled only for a finite set of points and in a finite imaginary-time
interval, and its evaluation is affected by the statistical noise,
which we minimize by using a quite high statistics. This integral
equation~(\ref{eqn:InversionProblem}) belongs to a class of
numerically ill-posed problems. Therefore, the use of special
techniques is warranted in order to extract numerically stable
results. We have employed an approach which combines two complementary
methods: the Singular Value Decomposition (SVD) and the
self-consistent Maximum Entropy Method (MEM), both described in great
detail in Ref.~\cite{MagierskiWlazlowski} and in the Supplemental
Material of Refs.~\cite{Wlazlowskietal,Wlazlowskietal2}.  The
stabilization procedure requires \apriori information about the
solution $\eta(\omega)$. The \apriori information used is related to
the known properties: the non-negativity of the shear
viscosity $\eta(\omega)\geqslant 0$, the asymptotic tail behavior
$\eta(\omega\to\infty)=\frac{C}{15\pi\sqrt{\omega}}$, and the sum rule
(see~\cite{TaylorRanderia} with subsequent
corrections~\cite{Enssetal,Hofmann})
%%%%%%%%%%%%%%%%%%%%%%%%%%%%%%
\begin{equation}
 \dfrac{1}{\pi}\int_{0}^{\infty}d\omega\left[ \eta(\omega)
 - \dfrac{C}{15\pi\sqrt{\omega}}\right] = \dfrac{\varepsilon}{3}-\dfrac{C}{12\pi a}, 
 \label{eqn:sumrule}
\end{equation}  
%%%%%%%%%%%%%%%%%%%%%%%%%%%%%%
where $C$ is Tan's contact density~\cite{Tan1}, and $\varepsilon$ is
the energy density. Both quantities are obtained within the same QMC
simulation.  The contact density $C$ was extracted from the analysis
of the tail of the numerically generated momentum distribution, which
for sufficiently large momenta decays as $n(p) \sim \frac{C}{p^{4}}$;
a similar technique was used in Ref.~\cite{Drutetal2} and subsequent
studies. Moreover, self-consistent MEM requires an appropriately
chosen class of \apriori models for the solution. Based on our past
experience for the unitary
limit~\cite{Wlazlowskietal,Wlazlowskietal3}, we determined that the
expected suitable models for the frequency dependent shear viscosity
$\eta(\omega)$ consist of Lorentzian-like structures at low
frequencies, smoothly evolving into the asymptotic tail behavior:
%%%%%%%%%%%%%%%%%%%%%%%%%%%%%%
\begin{eqnarray}
 M(\omega,\{\mu,\gamma,c,\alpha^{}_{1},\alpha^{}_{2}\}) =
 f(\omega,\{\alpha^{}_{1},\alpha^{}_{2}\})\,\dfrac{C}{15\pi\sqrt{\omega}}\nonumber\\
 +[1\!-\!f(\omega,\{\alpha^{}_{1},\alpha^{}_{2}\})]\, \mathcal{L}(\omega,\{\mu,\gamma,c\}),
\end{eqnarray} 
where
\begin{equation}
  f(\omega,\{\alpha^{}_{1},\alpha^{}_{2}\})=
\exp(-\alpha^{}_{1}\alpha^{}_{2})
\dfrac{\exp(\alpha^{}_{1}\omega)-1}{1+\exp(\alpha_1(\omega-\alpha_2))}
\end{equation} 
%%%%%%%%%%%%%%%%%%%%%%%%%%%%%%
and
\begin{equation}
 \mathcal{L}(\omega,\{\mu,\gamma,c\})=c\,\dfrac{1}{\pi} \, \frac{\gamma}{(\omega-\mu)^2 + \gamma^{2}}.
 \label{eqn:Lorentzian}
\end{equation}
%%%%%%%%%%%%%%%%%%%%%%%%%%%%%%
The parameters $\{\mu,\gamma,c,\alpha^{}_{1},\alpha^{}_{2}\}$ describe
admissible degrees of freedom of the model and are adjusted
automatically in a self-consistent manner.

\section{III. RESULTS}

\begin{figure}
\includegraphics[width=\pictsize\columnwidth]{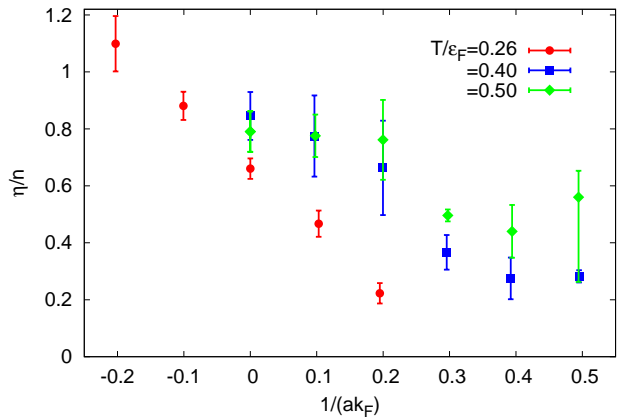}
\caption{ (Color online) The dimensionless static shear viscosity
$\frac{\eta}{n}$ as a function of interaction strength $\frac{1}{a\pF}$ for a $10^3$
lattice for selected temperatures: $T=0.26\eF$ - solid (red) circles,
$T=0.4\eF$ - (blue) squares and $T=0.5\eF$ - (green) diamonds.
The error bars show contribution to the computation uncertainty
generated by the analytic continuation procedure only. 
\label{fig:eta_over_n} }
\end{figure}

In~\fig{fig:eta_over_n}, the dimensionless static shear viscosity
$\frac{\eta}{n}$ is shown for three selected temperatures
$\frac{T}{\eF}=0.26$, $0.4$ and $0.5$ as a function of interaction
strength $1/(a\pF)$.  Only one point, for $\frac{T}{\eF}=0.26$ and
$\frac{1}{a\pF}=0.2$, corresponds to the system being in the
superfluid phase, while all other points correspond to the system
either in the normal phase or in the ``pseudogap'' regime.  We
emphasize that our QMC simulations are fully consistent with the existence
of the ``pseudogap'' regime, i.e.,  a temperature regime above the critical
temperature $T_c$ where many Cooper pairs are present, even though the
superfluidity is lost~\cite{Wlazlowskietal2,MWB}.  The presence of
pairs above the critical temperature is a property of the UFG well
established in \textit{ab initio} calculations. Naturally, this is also beyond
controversy in the BEC limit as well, where the critical temperature
is well below the dimer binding energy,
$T_c \ll \frac{\hbar^2}{ma^2}$.  The presence of a pseudogap regime is
to some extent also confirmed by experiments~\cite{Stewart:2008,
  Gaebler:2010, Sagi:2015}.  The present QMC results clearly show that
the static shear viscosity decreases as we enter into BEC regime.  As
we tune the system towards the BEC side of the Feshbach resonance the
number of Monte Carlo samples required to get an acceptable
signal-to-noise ratio increases rapidly.  For this reason we were not
able to perform reliable inversions of the
Eq. \eqref{eqn:InversionProblem} for interaction strengths beyond
$\frac{1}{a\pF}>0.2$ for temperature $\frac{T}{\eF}=0.26$ and beyond
$\frac{1}{a\pF}>0.5$ for temperatures $\frac{T}{\eF}=0.4$ and
$0.5$. The ratio of the shear viscosity to the number density
$\frac{\eta}{n}$ appears to have a relatively weak temperature
dependence above $T_c$.  This behavior appears to be confirmed by the
experiment~\cite{Elliott} as well.  Our results for temperatures
$\frac{T}{\eF}=0.4$ and $0.5$ suggest that there exists a minimum for
the static shear viscosity located on the BEC side of resonance for an
interaction strength corresponding to $\frac{1}{a\pF}\approx 0.4$.
However, the present QMC uncertainties do not permit an accurate
determination of the minimum position.

%%%%%%%%%%%%%%%%%%%%%%%%%%%%%%
%%%%%%%%%%%%%%%%%%%%%%%%%%%%%%

Comparing our results with experimental data~\cite{Elliott} we find
qualitative agreement. While the experiment reports existence of
minimum for the shear viscosity for $\frac{1}{a\pF}\simeq 0.25$, the
\textit{ab initio} prediction sets it for slightly higher
strengths. In general, the minimum position can be temperature
dependent, but experimental data suggest that this dependence is
rather weak, in apparent agreement with our findings. Note, however,
that the measurements are performed in trapped systems and only the
trap-averaged viscosity
$\avg{\frac{\eta}{n}}=\frac{1}{N\hbar}\int \eta(\bm{r})\,d^{3}\bm{r}$
is reported in Ref.~\cite{Elliott}.  The reduced temperature
$\frac{T}{\eF}$ as well as the interaction strength $\frac{1}{a\pF}$
are position dependent in a trap and they diverge to infinity as one
approaches the trap edges, where the gas enters the collisionless
regime and where the hydrodynamical approach is
inapplicable~\cite{Urban_2015}.  The trapped experiment probes all
reduced temperatures and interaction strengths, starting from the
values at the center of the cloud up to very high values. Moreover,
the regions further from the center of the trap also contribute with
increasingly higher weights in such averaged quantities, masking to a
large extent the information about the inner regions of the trap and
making its evaluation very challenging. To what extent the averaging
procedure affects the results for shear viscosity is not clear at
this time, and since a reliable validation of the theoretical
predictions against the experimental data requires knowledge of
the shear viscosity for all temperatures and interaction strengths,
such a comparison is beyond the scope of the present work. A similar
disagreement with experimental values was noted in the analysis
performed by Bluhm and Sch\"{a}fer~\cite{schaefer_2014}.  For example,
at unitarity the values extracted in the experimental analysis of
Ref.~\cite{Joseph} of the ratio $\frac{\eta}{n}$ exceed unity at
temperatures $\frac{T}{\eF} > 0.3$. The calculated ratio
$\frac{\eta}{n}$ within the kinetic theory~\cite{schaefer_2014}
attains such values only for significantly larger temperatures
$\frac{T}{\eF} \approx 1$ at unitarity (see Fig. \ref {fig:expcmp})
and even greater temperatures for positive values of the scattering
length.  The kinetic theory is appropriate for temperatures above
$T^*$, where pairs have completely dissociated and where the collision
integral is more or less well defined.  Below $T^*$ a more complex
kinetic approach is required, which should include dimer-dimer,
fermion-fermion, and dimer-fermion collisions as well as a
dimer-to-two-fermions and its time-reverse processes.  On the other
hand, when extrapolated, the kinetic theory and the present QMC
results appear, surprisingly to some extent, to be in agreement.

In order to confront the QMC results with the KSS conjuncture one has
to have information about the entropy density $s=\frac{S}{V}$. This can be
extracted from static observables including the energy $E$, the
chemical potential $\mu$, and the contact $C$, which are easily
obtained within the QMC framework. Combining the basic thermodynamic
relation (where $T$, $P$, and $N$ are, respectively, temperature,
pressure, and particle number)

%%%%%%%%%%%%%%%%%%%%%%%%%%%%%%
\begin{equation}
 E=TS - PV + \mu N
\end{equation} 
%%%%%%%%%%%%%%%%%%%%%%%%%%%%%%
together with Tan's pressure relation~\cite{Tan2}
%%%%%%%%%%%%%%%%%%%%%%%%%%%%%%
\begin{equation}
 P-\dfrac{2}{3}\dfrac{E}{V} = \dfrac{C}{12\pi a V},
\end{equation} 
%%%%%%%%%%%%%%%%%%%%%%%%%%%%%%
one can show that
%%%%%%%%%%%%%%%%%%%%%%%%%%%%%%
\begin{equation}
 \dfrac{S(x,y)}{N} = \dfrac{\xi(x,y)-\zeta(x,y)+\frac{1}{6\pi}\tilde{C}(x,y)y}{x},
\end{equation} 
%%%%%%%%%%%%%%%%%%%%%%%%%%%%%%
where we introduced the following dimensionless quantities: the
reduced temperature $x=\frac{T}{\eF}$, the strength of the interaction
$y=(a\pF)^{-1}$, the Bertsch parameter $\xi=\frac{5E}{3N\eF}$, the
reduced chemical potential $\zeta=\frac{\mu}{\eF}$, and the reduced contact
parameter $\tilde{C}=\frac{C}{N\pF}$.

\begin{figure}
\includegraphics[width=\pictsize\columnwidth]{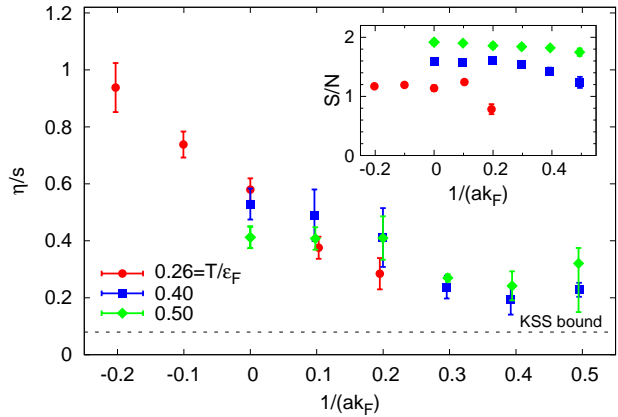}
\caption{ (Color online) The ratio of the shear viscosity to entropy
density $\frac{\eta}{s}$ as a function of interaction strength $\frac{1}{a\pF}$ for a
$10^3$ lattice.  The notation for theoretical series is identical
to~\fig{fig:eta_over_n}.  The black dashed line indicates KSS bound:
$\frac{1}{4\pi}$. The error bars encapsulate uncertainty originating from 
the analytic continuation procedure and statistical uncertainty of 
the entropy density determination.
The inset shows corresponding values of entropy per
particle.
\label{fig:eta_over_s} }
\end{figure}

In~\fig{fig:eta_over_s} we show the ratio of the shear viscosity to
the entropy density $\frac{\eta}{s}$. These results suggest that there is
minimum for this ratio located at $\frac{1}{a\pF}\approx 0.4$. However, the
present QMC uncertainties do not permit one to make a very precise
determination of its location. Moreover, the QMC data also reveal a
rather weak temperature dependence, similar to experimental findings.

%%%%%%%%%%%%%%%%%%%%%%%%%%%%%%
%%%%%%%%%%%%%%%%%%%%%%%%%%%%%%

The $\frac{\eta}{s}$ ratio at the minimum
$\left . \frac{\eta}{s}\right |_{min} \approx 0.2$,
is about $2.5$ times smaller than its value at the unitary limit, and
only $2.5$ times larger than the KSS holographic bound
$\frac{1}{4\pi}$. In the inset we also provide the
entropy dependence on the coupling constant for our selected
temperatures. For the temperature $\frac{T}{\eF}=0.26$ the value of
the entropy for $\frac{1}{a\pF}=0.2$, which visually appears to
deviate from the smooth pattern, reflects the fact that there is a
phase transition from the normal to the superfluid state as we
increase interaction strength at a fixed temperature.

\section{IV. QUANTUM \& CLASSICAL TURBULENCE}

The viscosity is at the root of classical turbulence.  In an ideal
fluid turbulence does not exist, while it can develop in either a
normal fluid or in a superfluid.  In superfluids viscosity was
expected to play a minor role (only in the normal component) and there
was no clear route to turbulence when the temperature tends to zero
and the viscosity therefore decreases. However, in 1955
Feynman~\cite{feynman} conjectured that in superfluids the crossing
and reconnection of quantized vortices could lead to ``quantum
turbulence,'' a field which since then is one of the most active areas
of research in the physics of liquid helium 3 and 4~\cite{donnelly,
  vinen, paoletti}. Feyman's conjecture was demonstrated to be the
correct theoretical mechanism in a dilute Bose
superfluid~\cite{koplik} and in a dilute Fermi superfluid as
well~\cite{aurel,gabriel}.  The theoretical framework of classical
turbulence laid by Kolomogorov~\cite{kolmogorov} appears to explain
many features (though not all) of quantum turbulence in liquid
helium~\cite{vinen}. At the same time many characteristics, such as
the non-Gaussian velocity distributions, are drastically
different~\cite{paoletti} in classical and quantum turbulence. The
non-Gaussian velocity distributions have been predicted to emerge also
in the UFG~\cite{gabriel}. The quark-gluon plasma and the Fermi gas in
the unitary regime above the critical temperature are two physical
systems, which are not superfluid, but in which the the shear
viscosity attains extremely low values.  Quantized vortices do not
exist either in the UFG above $T_c$ or in the quark-gluon plasma and
the dynamics of these systems can be very close to that of a
hypothetical classical fluid with zero viscosity, called the ideal
fluid, if existence of classical turbulence is prohibited.  Indeed,
this is a case of ultracold fermionic gases produced experimentally.
Classical fluid hydrodynamics is governed by dimensionless numbers,
where the most important is the Reynolds number $Re=\frac{nmvL}{\eta}$
, $v$ and $L$ are characteristic velocity and linear dimension
describing flow.  Classical turbulence in three-dimensional systems is achieved for
values of the Reynolds number of the order of $10^4$.

In a UFG quantized vortices, and therefore quantum turbulence, can
exist in clouds with as little as 500-1000
fermions~\cite{aurel,gabriel} and for flow velocities
$v\approx 0.7v_F$, which are larger than the Landau's critical velocity
$v_c\approx 0.4v_F$.  The largest cold atomic clouds created so far in
the laboratory have ${\cal O}(10^6)$ atoms.

One can estimate the Reynolds number for a UFG using as characteristic
scales: number of atoms $N$ in a cloud that defines density $n=N/L^3$,
the critical velocity
$v_c=0.4v_F=0.4\frac{\hbar(3\pi^2N)^{1/3}}{mL}\approx 1.2 N^{1/3}\hbar/mL$, 
and the minimal value of the shear viscosity $\frac{\eta}{n\hbar}\approx 0.2$.
With this one obtains for the Reynolds number the values
$Re\lesssim\frac{n\hbar}{\eta}\frac{mLv_c}{\hbar} \approx 60$
for $N=1000$ and $Re\approx 620$ for $N=10^6$.  One
can argue that one can attain higher values of the Reynolds number by
increasing the flow velocity by a factor $\approx 10$ so as to reach
$Re \approx 10^4$. In a dilute Fermi gas near the unitary point the
scattering cross section is on average $\sigma \approx 4\pi/k_F^2$.  If the flow
velocity is increased to $v\approx 10v_F$, the cross section decreases
by a factor of $k^2/k_F^2\approx 100$ and the mean free path
$1/n\sigma$ becomes very large, comparable or exceeding the size of
any atomic cloud created so far in the laboratory and the system
enters the collisionless regime.  One could alternatively contemplate 
an increase in the
linear size of a cloud by a factor of 10, thus up to cloud particle
numbers ${\cal O}(10^9)$  (a size likely difficult to achieve for 
condensates), in order to increase the Reynolds number
by an order of magnitude.

The Fermi gas in the unitary regime is thus a rather unique physical
system; below the critical temperature the system is superfluid and
can sustain quantum turbulence in rather small clouds, while above the
critical temperature the turbulent dynamics is strongly suppressed for
any current experimental realizations for a very wide range of flow
velocities and cloud sizes.  The almost ``death'' of classical
turbulence above $T_c$ and its revival into a new ``body,'' the
quantum turbulence below $T_c$, makes the unitary Fermi gas the
unitary regime a quite unique physical system. The small value of the
shear viscosity in a dilute Fermi gas near unitarity, which is
attained in the normal phase, requires that the onset of classical
turbulence be achieved in relatively large clouds, so far not realized
experimentally.

In summary, we have presented \textit{ab initio} results for
interaction strength dependence of the static shear viscosity and the
shear viscosity to the entropy density ratio. Both quantities decrease
as we tune interaction strength from the BCS side of the unitarity
point towards the BEC limit. The results suggest that the Fermi gas in
the unitary regime is the closest physical system known to being a
``perfect fluid," for an interaction strength corresponding to
$\frac{1}{a\pF}\approx 0.4$, with a minimum value of shear viscosity
to entropy density
$\left . \frac{\eta}{s}\right |_{min} \approx 0.2 \frac{\hbar}{k_B}$
about twice as small than the value reported for the quark-gluon
plasma
$\left . \frac{\eta}{s}\right |_{QGP} \lesssim
0.4\frac{\hbar}{k_B}$~\cite{Song}.
Our simulations qualitatively confirm the experimental observation of
Ref.~\cite{Elliott}, that shear viscosity attains a minimum on the BEC
side of the unitary point, albeit for a stronger value of the coupling
constant.

This work was supported in part by U.S. Department of Energy (DOE)
Grant No. DE-FG02-97ER41014 and the Polish National Science Center
(NCN) under Contracts No. UMO-2013/08/A/ST3/00708 and
No. UMO-2014/13/D/ST3/01940.  Calculations reported here have been
performed at the University of Washington Hyak cluster funded by the
NSF MRI Grant No. PHY-0922770.

%%%%%%%%%%%%%%%%%%%%%%%%%%%%%%%%%%%%%%%%%%%%%%%%%%%

\end{document}